\pdfoutput=1

\documentclass[unnumsec,webpdf,contemporary,large]{oup-authoring-template}%

\usepackage{booktabs}
\usepackage{adjustbox}
\usepackage{tcolorbox} 
\graphicspath{{Fig/}}

\theoremstyle{thmstyleone}%
\theoremstyle{thmstyletwo}%
\theoremstyle{thmstylethree}%

\begin{document}

\journaltitle{Journal Title Here}
\DOI{DOI HERE}
\copyrightyear{2022}
\pubyear{2019}
\access{Advance Access Publication Date: Day Month Year}
\appnotes{Paper}

\firstpage{1}


\title[Short Article Title]{M$^{3}$-20M: A Large-Scale Multi-Modal Molecule Dataset for AI-driven Drug Design and Discovery}

\author[1]{Siyuan Guo\ORCID{0000-0003-0378-830X}}
\author[1]{Lexuan Wang\ORCID{0009-0009-8778-0217}}
\author[1]{Chang Jin\ORCID{0000-0003-0378-830X}}
\author[2]{Jinxian Wang}
\author[1]{Han Peng\ORCID{0000-0001-6519-8538}}
\author[1]{Huayang Shi\ORCID{0009-0007-0870-6832}}
\author[1]{Wengen Li\ORCID{0000-0002-8768-6740}}
\author[1,$\ast$]{Jihong Guan\ORCID{0000-0003-2313-7635}}
\author[2,$\ast$]{Shuigeng Zhou\ORCID{0000-0002-1949-2768}}

\authormark{Author Name et al.}

\address[1]{\orgdiv{Department of Computer Science and Technology}, \orgname{Tongji University}, \orgaddress{\street{No. 4800 Cao'an Road}, \postcode{201804}, \state{Shanghai}, \country{China}}}
\address[2]{\orgdiv{Shanghai Key Lab of Intelligent Information Processing, and School of Computer Science}, \orgname{Fudan University}, \orgaddress{\street{2005 Songhu Road}, \postcode{200438}, \state{Shanghai}, \country{China}}}

\corresp[$\ast$]{Corresponding author. \href{mailto:jhguan@tongji.edu.cn}{jhguan@tongji.edu.cn}, \href{mailto:sgzhou@fudan.edu.cn}{sgzhou@fudan.edu.cn}}

\received{Date}{0}{Year}
\revised{Date}{0}{Year}
\accepted{Date}{0}{Year}



\abstract{This paper introduces M$^{3}$-20M, a large-scale Multi-Modal Molecule dataset that contains over 20 million molecules, with the data mainly being integrated from existing databases and partially generated by large language models. Designed to support AI-driven drug design and discovery, M$^{3}$-20M is 71 times more in the number of molecules than the largest existing dataset, providing an unprecedented scale that can highly benefit the training or fine-tuning of models, including large language models for drug design and discovery tasks. This dataset integrates one-dimensional SMILES, two-dimensional molecular graphs, three-dimensional molecular structures, physicochemical properties, and textual descriptions collected through web crawling and generated using GPT-3.5, offering a comprehensive view of each molecule. To demonstrate the power of M$^{3}$-20M in drug design and discovery, we conduct extensive experiments on two key tasks: molecule generation and molecular property prediction, using large language models including GLM4, GPT-3.5, GPT-4, and Llama3-8b. Our experimental results show that M$^{3}$-20M can significantly boost model performance in both tasks. Specifically, it enables the models to generate more diverse and valid molecular structures and achieve higher property prediction accuracy than existing single-modal datasets, which validates the value and potential of M$^{3}$-20M in supporting AI-driven drug design and discovery. The dataset is available at \url{https://github.com/bz99bz/M-3}.}


\maketitle

\section{Introduction}
As an important field in pharmacy, drug design and discovery aim to identify novel therapeutic compounds and optimize their properties for clinical application~\cite{santos2017comprehensive,hughes2011principles,tsaioun2016evidence,wu2018moleculenet,chen2023artificial}. This process involves a series of tasks, from the prediction of molecule-target interactions~\cite{ji2023comprehensive,ye2021unified,li2022effective,lu2024dynamicbind} and molecular properties~\cite{fu2024admetlab,goldman2023prefix,rong2020self,masters2022gps++}, to the design or generation of targeted molecules~\cite{gao2022sample,franke2022probabilistic,kong2022molecule,yang2021hit,song2024equivariant}, which are fundamental for developing effective and safe drugs. In recent years, various models trained or pre-trained and fine-tuned using massive data, such as~\cite{bagal2021molgpt,chithrananda2020chemberta,honda2019smiles,fabian2020molecular,liu2024git,guo2023indeed,liu2023group,ai2024extracting,yu2023multimodal,ye2025drugassist}, have emerged as powerful tools in this domain. These models leverage large datasets and sophisticated algorithms to generate new compounds, predict molecular structures, and analyze their biochemical properties, significantly accelerating drug design and discovery. 
However, most existing molecule datasets for model training are limited to single-modality, failing to capture molecular characteristics comprehensively, thus hindering the development of powerful models for drug design and discovery

Recent works~\cite{liu2024git,liu2023multi,zeng2022deep} are paying increasing attention to multi-modal molecule datasets. Zeng~\cite{zeng2022deep} introduced a deep-learning system that bridges molecular structures and biomedical texts, achieving comprehension capacity comparable to human professionals. Liu et al.~\cite{liu2023multi} developed a multi-modal model for text-based retrieval and editing of molecular structures, facilitating advanced molecular text analysis. Liu et al.~\cite{liu2024git} presented a multi-modal large language model for molecular science that integrates graphs, images, and texts to support various downstream tasks. However, datasets used in these works still suffer from significant limitations. On the one hand, these datasets are typically small in size, encompassing a limited chemical space that restricts the generalization power of the trained or tuned models. On the other hand, the absence of complete modalities of molecular data undermines the performance of trained models. 

\begin{figure*}[!t]
\centering
{\includegraphics[width=1\textwidth]{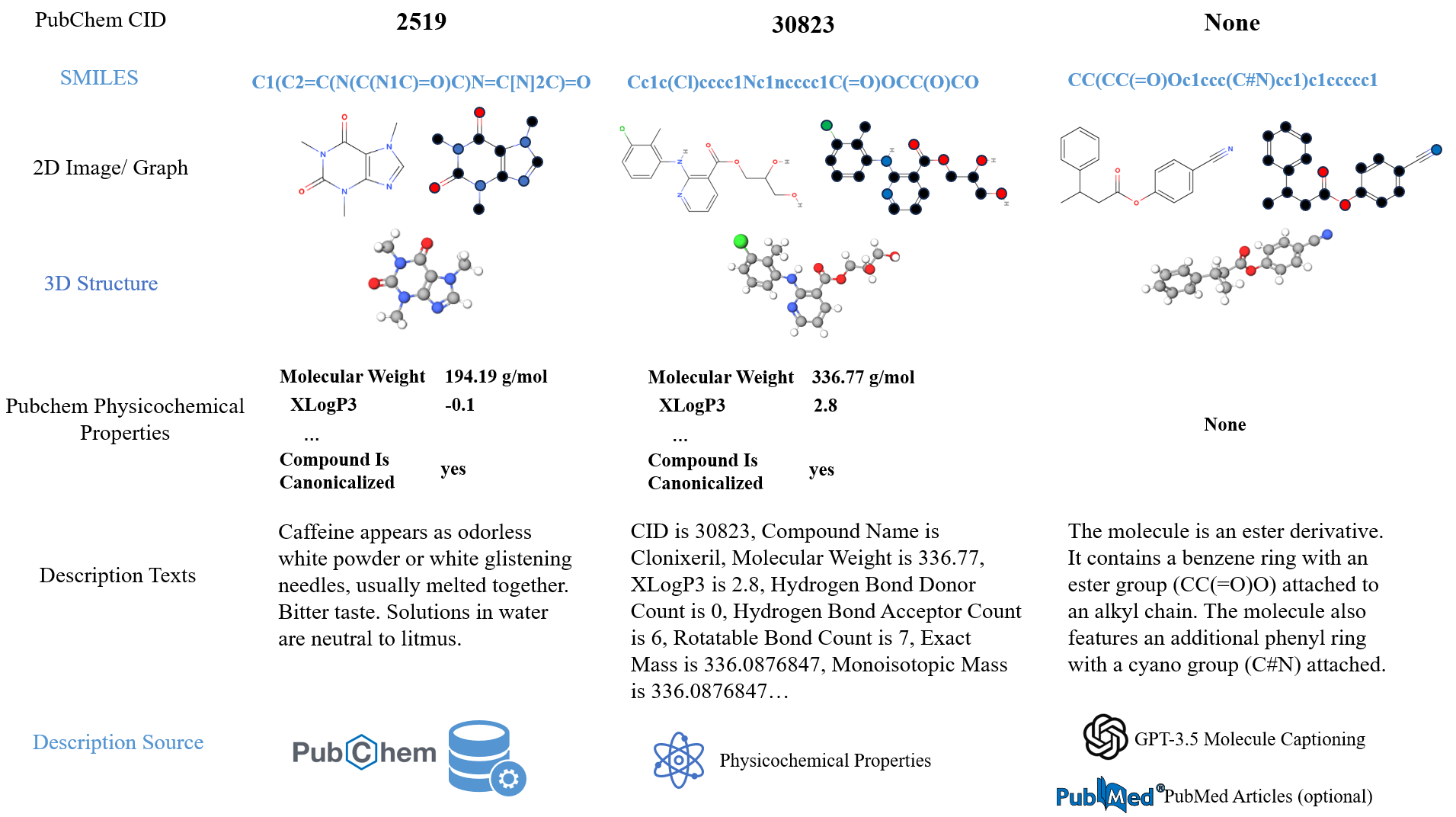}}
\caption{Some examples from our M$^{3}$-20M dataset.}
\label{fig:datasets}
\end{figure*}

To address these limitations, this paper presents a new and large-scale multi-modal integrated molecule dataset called M$^3$-20M for AI-driven drug design and discovery. Data in M$^3$-20M are mainly integrated from existing molecular databases, and supplemented by generation with large language models (LLMs).  M$^3$-20M contains more than 20 million molecules with their SMILES strings, 2D graphs, 3D structures, physicochemical properties, and textual descriptions, providing unprecedented data scale, diversity, and comprehensiveness for training models of drug design and discovery downstream tasks. M$^3$-20M is 71 times more in the number of molecules than the largest existing multi-modal dataset PubChemSTM~\cite{liu2023multi}. It collects physicochemical properties from the PubChem database~\cite{kim2016pubchem} and enriches the description texts of molecules by web crawling and GPT-3.5 generation. Therefore, each molecule's SMILES corresponds to its 2D graph and 3D structure, physicochemical properties, and description texts. Figure~\ref{fig:datasets} shows some examples from our M$^{3}$-20M dataset.
%

In summary, M$^3$-20M stands out of the existing datasets in at least the following three aspects: 1) \textbf{Large scale}: it contains over 20 million molecules, which is the largest open-access multi-modal molecule dataset for AI-driven drug design and discovery, to the best of our knowledge. %
2) \textbf{Comprehensive modalities}: 
M$^3$-20M boasts a more complete range of modalities, including one-dimensional molecular SMILES strings, two-dimensional molecular graphs, three-dimensional molecular structures, physicochemical properties, and textual descriptions. Such multi-modal data offers a holistic perspective of each molecule, benefiting model training and tuning for drug design and discovery. Additionally, we offer tools to generate 2D molecular graph images that facilitate compound patent retrieval, and to crawl PubMed for research articles related to specific molecules, thereby enriching textual descriptions.
3) \textbf{Supporting various tasks}: M$^3$-20M can support model training and tuning for various downstream drug design and discovery tasks, including molecule generation, molecular property prediction, lead optimization, and virtual screening. It also can aid in tasks such as pharmacokinetics modeling and drug-target interaction prediction. Moreover, we construct seven multi-modal sub-datasets, QM9-MM, MOSES-MM, BACE-MM, BBBP-MM, HIV-MM, ClinTox-MM, and Tox21-MM, to facilitate more accurate and robust molecular property prediction. 

To demonstrate M$^{3}$-20M's value and potential in supporting drug design and discovery, we conduct extensive experiments on two fundamental tasks,  molecule generation and molecular property prediction. Our evaluation encompasses three categories of models, including closed-source LLMs (GLM4~\cite{du2022glm}, GPT-3.5~\cite{ouyang2022training}, and GPT-4~\cite{achiam2023gpt}), open-source LLM (Llama3-8b~\cite{dubey2024llama}) and specialized molecule generation models (MoFlow~\cite{zang2020moflow} and D2L-OMP~\cite{guo2024diffusing}). 
Our experimental results show that M$^{3}$-20M can significantly boost model performance in both tasks. Concretely, with M$^{3}$-20M, the above models can generate more diverse and valid molecular structures, and achieve higher property prediction accuracy than using existing single-modal datasets. 

\section{Related work}\label{sec2}

Multi-modal molecular data offers comprehensive AI and molecular biology insights, enhancing drug discovery, property prediction, and generation tasks. As Wang et al.~\cite{wang2023multitask} demonstrated, integrating sequence, structure, and text data improves drug discovery accuracy. However, challenges persist in organization, processing, and maintaining consistency across different modalities. There's an urgent need for large-scale integrated multi-modal datasets for drug design model development. In what follows, we briefly review the related work in three perspectives, molecular representation, single and two-modality molecule datasets, and multi-modality molecule datasets. 

\subsection{Molecular representations}  
Molecules can be represented through diverse modalities~\cite{david2020molecular,ertl2010molecular}: 1D sequences (SMILES, fingerprints, topological characterizations), 2D visualizations (graphs, RDKit/Open Babel~\cite{landrum2013rdkit,o2011open} images), 3D structures (atomic coordinate graphs/grids~\cite{li2022deep}), and textual descriptions from experiments. Single-modal data is a reduction of the complete structural information of the molecules, which means the representations learned from single-modal suffer from information loss. For example, two SMILES may represent the same molecule or a two-dimensional molecular graph cannot distinguish between the conformations. Therefore, integrating multi-modal information is crucial for molecular modeling and model training in drug design and discovery. 

\subsection{Single-modal and two-modal molecule datasets}

Existing molecule datasets are predominantly limited to single or dual modalities (1D SMILES strings, 2D structures, or 3D conformations), restricting the generalization power of trained models~\cite{wang2023multimodal,fink2007virtual,wang2022deep,axelrod2022geom,yang2024quandb}. For instance, MOSES~\cite{polykovskiy2020molecular} comprises 4,591,276 molecules with only SMILES representations, 
while ZINC~\cite{irwin2005zinc} includes over 1.3B molecules with 2D graphs and some 3D structures. 
Although recent advances like Zeng et al.~\cite{zeng2022deep} presented a molecular structure-text system. Beaini et al.~\cite{beaini2023towards} developed Graphium, a vast repository of 100M molecules with 13B labels, yet it focuses on multi-task learning rather than multi-modal representation. 

\subsection{Multi-modal molecule datasets} 

The growth of AI-driven drug design and discovery has elevated the importance of multi-modal molecular data.
Current multi-modal datasets remain limited: PubChemSTM~\cite{liu2023multi} contains 280,000 structure-text pairs but lacks sufficient volume; igcdata~\cite{liu2024git} includes 220,000 descriptions from PubChem and ChEBI-20~\cite{gaulton2012chembl}; and ChEBI-20-MM~\cite{liu2025quantitative} provides 32,998 molecules as an evaluation benchmark rather than a training dataset. These datasets suffer from inadequate scale, limited modality coverage, and sparse textual descriptions.


To overcome these limitations, we present the M$^{3}$-20M dataset, which significantly expands the scale and scope of multi-modal molecular data. Table~\ref{tab:multimodal_datasets} compares our dataset with existing multi-modal datasets across seven dimensions: molecule count, 3D structures, 2D graphs, PubChem CID, physicochemical properties, downstream tasks, and description texts. M$^3$-20M has the largest number of molecules and covers the most comprehensive molecular data modalities.

\begin{table*}[t]
\centering
\caption{Comparison between our dataset M$^{3}$-20M and existing multi-modal datasets.}
\label{tab:multimodal_datasets}
\begin{tabular}{>{\centering\arraybackslash}m{2.6cm} >{\centering\arraybackslash}m{2cm} >{\centering\arraybackslash}m{2cm} >{\centering\arraybackslash}m{3cm} >{\centering\arraybackslash}m{2.3cm} >{\centering\arraybackslash}m{2cm}}
\toprule
\textbf{Property} & \textbf{QM9\cite{ramakrishnan2014quantum}} & \textbf{PCdes\cite{zeng2022deep}} & \textbf{PubChemSTM\cite{liu2023multi}} & \textbf{igcdata\cite{liu2024git}} & \textbf{M$^3$-20M (Ours)} \\
\midrule
\textbf{SMILES} & 134K & 1.5K & 280K & 220K & 20M \\
\textbf{3D structures} & $\checkmark$ & × & × & × & $\checkmark$ \\
\textbf{2D graphs} & $\times$ & $\times$ & $\checkmark$ & × & $\checkmark$ \\
\textbf{2D images} & $\times$ & $\times$ & × & $\checkmark$ & $\checkmark$ \\
\textbf{PubChem CID} & $\times$ & × & $\checkmark$ & × & $\checkmark$ \\
\textbf{Physicochemical properties} & $\times$ & × & × & × & $\checkmark$ \\
\textbf{Downstream task datasets} & $\times$ & × & × & × & $\checkmark$ \\
\textbf{\#Text descriptions} & $\times$ & 1.5K & 280K & 220K & 20M \\
\bottomrule
\end{tabular}%
\end{table*}

\section{Methodology}
This section outlines our dataset construction process and statistics. Molecular data is collected from existing databases PubChem, ZINC, and QM9\cite{ruddigkeit2012enumeration,ramakrishnan2014quantum}, processing them into multiple representations of molecules, including SMILES strings, 2D and 3D structures, physicochemical properties, and textual descriptions. Notably, we enhance molecular descriptions using GPT-3.5 with an expert scoring mechanism to ensure scientific accuracy. The final dataset integrates over 20 million molecules.

\begin{figure*}[!t]
\centering
\includegraphics[width=1\textwidth]{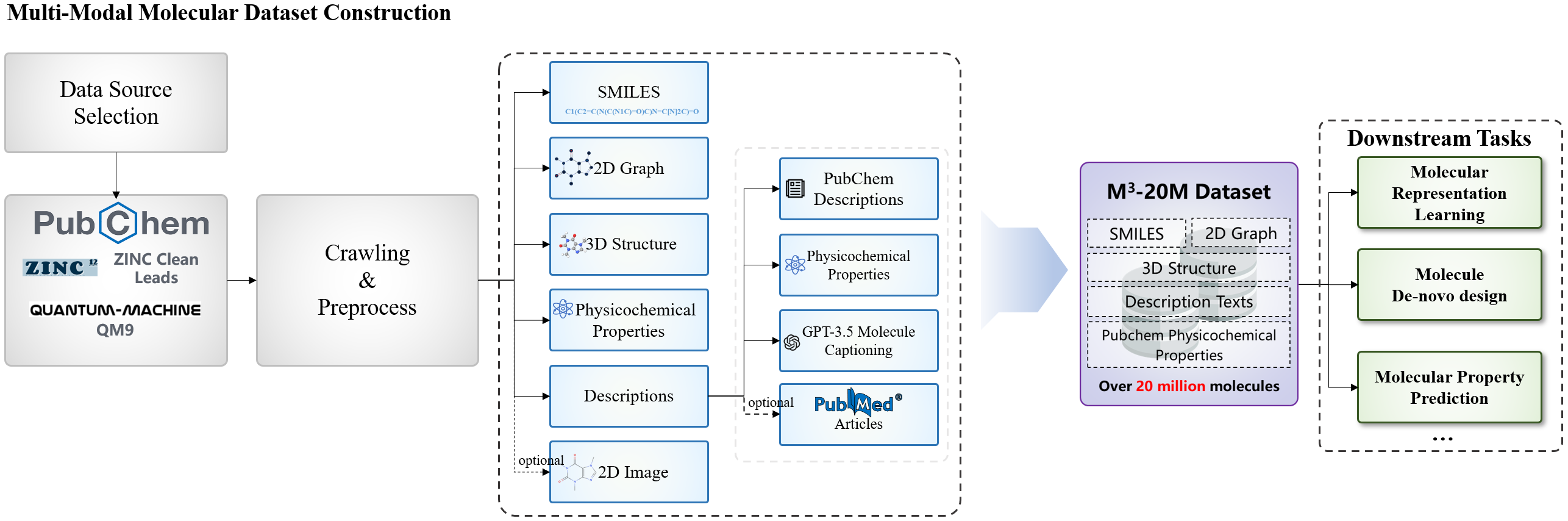}
\caption{The construction pipeline of our multi-modal molecule dataset M$^{3}$-20M.}
\label{fig:data_construction}
\end{figure*}

\subsection{Data Collection}
Figure~\ref{fig:data_construction} illustrates our M$^{3}$-20M construction pipeline. To obtain molecules and the corresponding molecular graphs and 3D structures, we first collect molecules from the PubChem, ZINC, and QM9 databases for their extensive molecular coverage and reliable data quality. Then, we utilize the Chem function of RDKit to extract atomic features and chemical bond characteristics and convert them into 2D molecular graphs. Additionally, we employ the PubChem API to batch download SDF files containing the 3D structures of molecules and use the GetAtomPosition function of RDKit to calculate the 3D coordinates of the central atoms.

Textual descriptions serve as a bridge for large language models to comprehend multi-modal molecules. For textual descriptions, we employ three approaches: 
1) For the molecules that have textual descriptions in PubChem, we directly extract the texts from PubChem. 2) For the molecules that do not have textual descriptions but have physicochemical properties in PubChem, we transform these physicochemical property values into textual descriptions, which will be detailed later. 3) For those without physicochemical property values or not present in the PubChem database, we generate their textual descriptions by GPT-3.5. We present the details in the following sections. 

Initially, we utilized textual data from PubChem; however, only a subset of molecules in PubChem contains associated textual descriptions. The sparsity of text information leads to poor performance of the trained or tuned models. So we resort to text augmentation. To generate textual descriptions for molecules having physicochemical property values in PubChem, we extract 26 key physicochemical properties and their values from PubChem (Table~\ref{tab:chemical_physical_properties}).

Concretely, we search for these properties using the compound identifier (CID) in PubChem, ensuring all data complied with licensing requirements and utilizing the open interface provided by PubChem. After collecting 19,175,245 pieces of raw data, we transform the properties of each molecule into structured descriptions using the template ``\textit{property name} is a \textit{specific value}''. Here, ``\textit{property name}'' indicates any of the 26 properties, and ``\textit{specific value}'' means a concrete value of the property.

Additionally, we expand seven downstream task datasets (MOSES, QM9, BBBP, BACE, HIV, Tox21, ClinTox) with multi-modal information and the PubChem CIDs, creating their "-MM" counterparts to enhance both molecule generation and property prediction tasks.


\subsection{Text Description Generation by GPT-3.5}

\begin{figure*}[!t]
\centering
\includegraphics[width=1\textwidth]{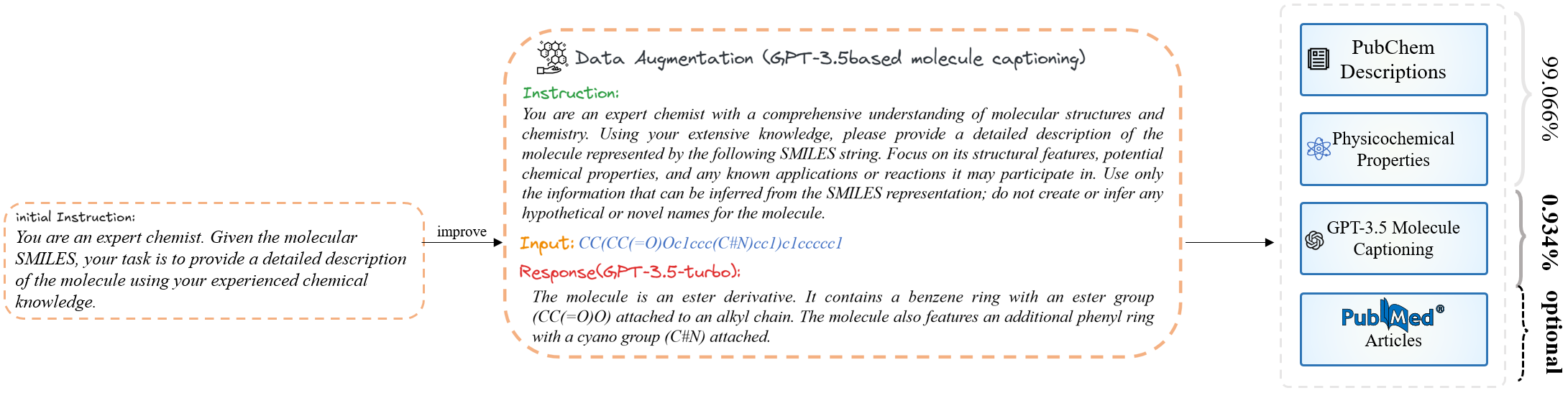}
\caption{Molecular text description generation by GPT-3.5.}
\label{fig:text_generation}
\end{figure*}

For molecules lacking physicochemical properties or not present in PubChem, we generate textual descriptions using GPT-3.5. Figure \ref{fig:text_generation} illustrates the process of molecular description text generation. Initially, we used a basic prompt identifying GPT-3.5 as an expert chemist, but this produced errors in SMILES interpretation and functional group identification.

Inspired by the CO-STAR (Context-Objective-Style-Tone-Audience-Response) framework, we develop an enhanced prompt: ``\textit{You are an expert chemist with a comprehensive understanding of molecular structures and chemistry. Using your extensive knowledge, please describe the molecule represented by the following SMILES string. Focus on its structural features, potential chemical properties, and any known applications or reactions it may participate in. Use only the information inferred from the SMILES representation; do not create or infer any hypothetical or novel names for the molecule}.''

Utilizing this enhanced prompt, we generated 1,073,845 descriptions, which is only 0.934\% of the total descriptions. The proportion of synthetic data to real data is very low and will not have a negative impact on real data. The subsequent section details our Human Expert Scoring Mechanism for quality control.

\subsection{Generated Text Quality Control by Human Expert Scoring}

To ensure high-quality textual descriptions, we design a quality control protocol inspired by some recent works~\cite{mccann2018natural,kwiatkowski2019natural}, which consists of three parts: selection and training of evaluation experts (or validators), scoring criteria, and stratified random sampling. 

\subsubsection{Evaluation Expert Selection and Training}
To evaluate the quality of generated texts, six chemistry specialists (3 undergraduate and 3 postgraduate students) from top Chinese universities with documented expertise in molecular characterization were recruited as validators. All validators received a standardized training program covering evaluation criteria, error classification, and scoring calibration using 200 pilot samples. 

\subsubsection{Quality Assessment Criteria}
This mechanism assesses the textual descriptions from four dimensions: \textit{accuracy}, \textit{effectiveness}, \textit{comprehensiveness}, and \textit{simplicity}, as established in previous works on natural language generation and evaluation~\cite{wang2018glue,post2018call}. Each description can earn a maximum score of 10 points. The scoring system is structured as follows:
\begin{enumerate}
\item \textbf{Accuracy (Maximum 5 points)}
\begin{enumerate}
\item \textbf{Correct Description (5 points):} The molecular description is entirely accurate, containing no scientific errors. This includes correct usage of chemical terms, accurate structural representation, and proper property identification.
\item \textbf{Slight Errors (2-4 points):} Minor inaccuracies are present, such as small structural mistakes or minor property deviations. These errors should not impede overall comprehension. Notably, any incorrect identification of functional groups results in a 1 point deduction.
\item \textbf{Serious Errors (0 points):} Major inaccuracies are present, such as entirely incorrect chemical terminology or wrong molecular names, stripping the description of scientific value.
\end{enumerate}

\item \textbf{Effectiveness (Maximum 2 points)}
\begin{enumerate}
\item \textbf{Highly Valid (2 points):} The description highlights key molecular properties, such as reactivity or pharmacodynamics, and may include useful applications or future prospects.
\item \textbf{Moderately Effective (1 point):} While the description includes valid information, it may lack detailed analysis or specific application scenarios.
\item \textbf{Ineffective Description (0 points):} The description fails to convey crucial properties or applications and lacks practical utility.
\end{enumerate}

\item \textbf{Comprehensiveness (Maximum 2 points)}
\begin{enumerate}
\item \textbf{Very Comprehensive (2 points):} The description extensively covers multiple aspects of the molecule, including structure, physicochemical properties, reactivity, and potential applications.
\item \textbf{Fairly Comprehensive (1 point):} Several aspects of the molecule are described, though some areas may lack depth.
\item \textbf{Incomplete (0 points):} The description is limited to one aspect of the molecule and significantly lacks breadth.
\end{enumerate}

\item \textbf{Simplicity (Maximum 1 point)}
\begin{enumerate}
\item \textbf{Concise (1 point):} Succinct, clearly articulated, and free of redundant information.
\item \textbf{Not Concise (0 points):} Verbose and repetitive, detracting from clarity and conciseness.
\end{enumerate}
\end{enumerate}

Each text description is evaluated based on a maximum score of 10 points. For every 100 generated molecules, 10\% are randomly selected for expert scoring. Descriptions scoring above 5 points are considered qualified, while those below 5 points are regenerated. Descriptions with significant inaccuracies (0 points in accuracy) will also be regenerated.

\subsubsection{Sampling Strategy and Validation Results}
We adopt stratified random sampling to sample 12,000 molecules (2,000 per validator) from our dataset, ensuring no overlap between validators. The validation reveals that the qualification rate is 98.8\% (11,856/12,000), i.e., with only 144 instances requiring regeneration (24 ± 3.1 per validator). All the regenerated text descriptions underwent a secondary validation. This rigorous validation ensures that $\geq$ 99\% of released descriptions meet strict scientific standards. 

\subsection{Multi-Modal Datasets for Downstream Tasks}
To support model training for downstream tasks, we construct multi-modal datasets by integrating PubChem CIDs and descriptive text, thereby providing richer semantic context and molecular identity. In particular, we derive seven additional datasets from M$^3$-20M, they are \textbf{MOSES-MM} and \textbf{QM9-MM} for molecule generation, while \textbf{QM9-MM}, \textbf{BBBP-MM}, \textbf{BACE-MM}, \textbf{HIV-MM}, \textbf{Tox21-MM}, and \textbf{ClinTox-MM} for molecular property prediction. Table~\ref{tab:dataset_downstream} summarizes these datasets.

\begin{table*}[h!]
\centering
\caption{Statistics of the multi-modal datasets this work augmented for downstream tasks, including the number of tasks, the number of molecules, task type, and evaluation metrics}
\label{tab:dataset_downstream}
\begin{tabular}{@{}lcccc@{}}
\toprule
\textbf{Dataset} & \textbf{\#Tasks} & \textbf{\#Molecules} & \textbf{Task type}       & \textbf{Metric} \\ 
\midrule
MOSES-MM           & 1                 & 1,936,962                 & Generation             & SNN/Test, Frag/Test, Scaf/Test, Filters, logP, SA, QED, weight             \\
QM9-MM             & 2                 & 6,830                 & Generation and Regression             & Validity, Uniqueness, Novelty, MAE             \\
BBBP-MM            & 1                 & 2,039                 & Classification          & ROC--AUC        \\
BACE-MM            & 1                 & 1,513                 & Classification          & ROC--AUC        \\
ClinTox-MM         & 2                 & 1,478                 & Classification          & ROC--AUC        \\
Tox21-MM           & 12                & 7,831                 & Classification          & ROC--AUC        \\
HIV-MM             & 1                 & 41,127               & Classification          & ROC--AUC        \\

\bottomrule
\end{tabular}
\end{table*}

\section{Experiments}
In this section, we evaluate the value and potential of the M$^3$-20M dataset by experiments of two fundamental downstream tasks: molecule generation and molecular property prediction. In the experiments, we employ two different model adaptation approaches:
(1) Few-shot prompting with in-context learning (FP-ICL)~\cite{brown2020language} on closed-source models (GLM4, GPT-3.5, and GPT-4), where we design input prompts with demonstration examples from M$^3$-20M to guide the models' responses without modifying model parameters;
(2) Parameter-efficient fine-tuning using low-rank adaptation of large language models (LoRA)~\cite{hu2021lora} on the open-source Llama3-8b model, which updates a small subset of model parameters while maintaining the base model's knowledge.
Additionally, we present comprehensive statistical analyses on M$^3$-20M to demonstrate its potential for supporting downstream drug design and discovery tasks. Our findings show that M$^3$-20M can significantly improve model performance in the two molecular tasks with both adaptation approaches.  

\subsection{LLM Adaptation Methods}
\begin{figure*}[!t]
\centering
\includegraphics[width=1\textwidth]{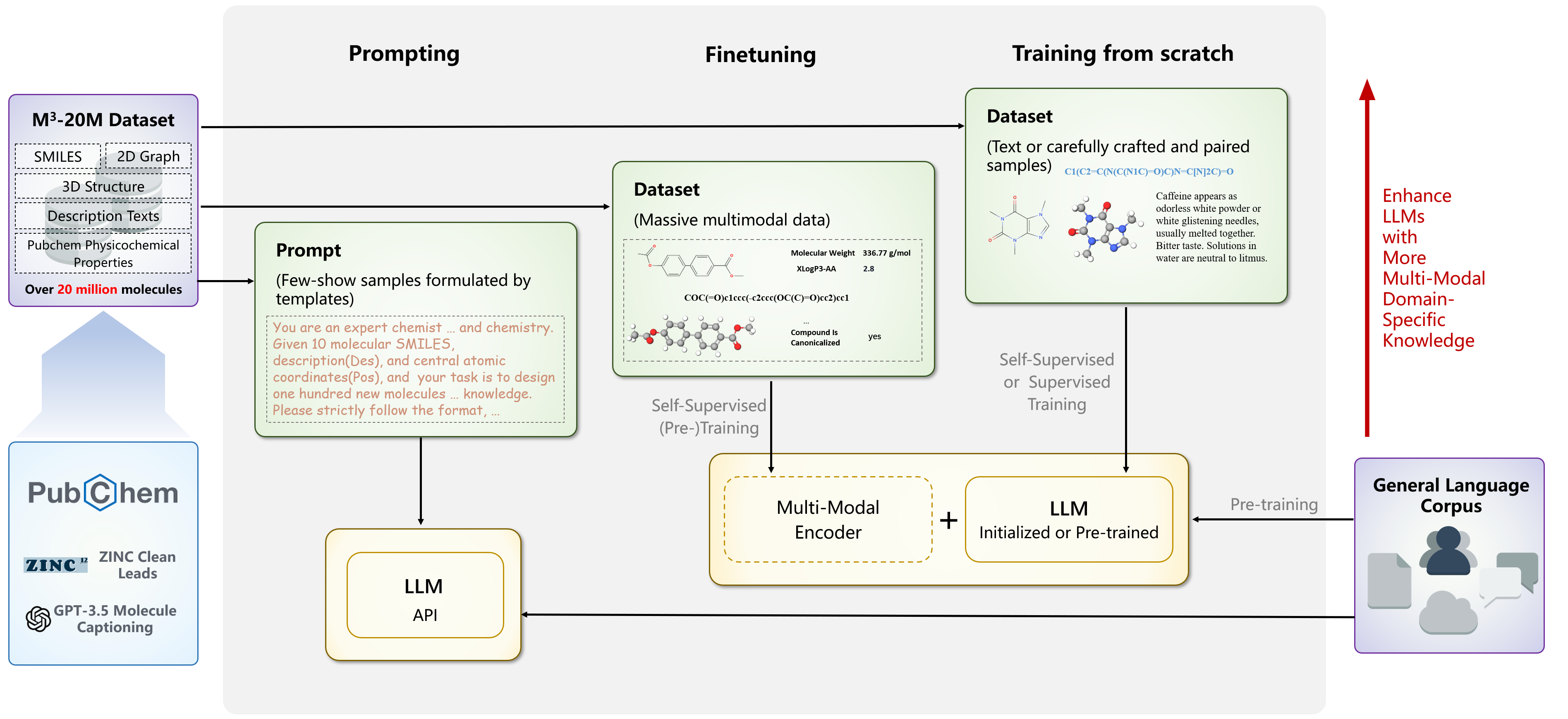}
\caption{Paradigms of model training and finetuning using our dataset.}
\label{fig:downstream}
\end{figure*}

Large language models (LLMs) can be adapted for domain-specific tasks through various methods. As illustrated in Fig.~\ref{fig:downstream}, we employ two methods that differ significantly in their resource requirements and implementation.

\subsubsection{Few-shot Prompting with In-context Learning (FP-ICL)} This approach involves providing carefully selected example pairs from M$^3$-20M alongside the target task in a single prompt. By incorporating these demonstrations, the model learns to recognize patterns and generate appropriate responses for new cases without any parameter update. This method requires only API access to LLMs (such as GPT-4, GPT-3.5, and GLM4) and minimal computational resources, making it highly accessible and cost-effective for practical applications.

\subsubsection{Parameter-efficient Fine-tuning with LoRA} Low-Rank Adaptation (LoRA) represents a more sophisticated adaptation method that modifies a subset of model parameters while preserving the model's base knowledge. This approach requires access to open-source models (such as Llama3-8b) and substantial computational resources (e.g. RTX4090 GPU). While more resource-intensive, LoRA enables deeper model adaptation for specialized tasks.

While M$^3$-20M is also suitable for training LLMs from scratch, such training requires powerful computational infrastructure beyond the scope of this study. Our adoption of these two adaptation methods provides practical insights into the dataset's utility under different resource constraints. 

\subsection{Molecule Generation Experiments}
The M$^3$-20M dataset enables de novo molecule generation by providing rich contextual information that improves the understanding of chemical structures and properties of generative models. 

\subsubsection{Experimental Details}
We employ the FP-ICL method to leverage multi-modal information for molecular generation by LLMs. Both general-purpose closed-source LLMs (GLM-4, GPT-3.5, GPT-4) and specialized generative models (MoFlow and D2L-OMP) are used. We compare the performance of using multi-modal molecules and single-modal molecules as examples for generation. For the multi-modal examples, we provide 10 randomly selected molecules from our M$^3$-20M dataset with their SMILES strings, 3D coordinates, and textual descriptions to the LLMs. The single-modal examples use only SMILES strings. 100 molecules per configuration are generated for performance assessment. We trained these models using QM9-MM to maintain compatibility with their existing network structures and parameters. Both models are trained on RTX4090 with 24GB memory and batch size is 32. The flow-based model (MoFlow) is trained for 200 epochs, while the diffusion-based model (D2L-OMP) is trained for 300 epochs. The rest of the parameters of the two models are the same as those in the original paper. We perform each experiment five times to ensure statistical reliability and report the averaged results and variance. 

We employ the FP-ICL method to leverage multi-modal information for molecular generation by LLMs. Both general-purpose closed-source LLMs (GLM-4, GPT-3.5, GPT-4) and specialized generative models (MoFlow and D2L-OMP) are used. We compare the performance of using multi-modal molecules and single-modal molecules as examples for generation. For the multi-modal examples, we provide 10 randomly selected molecules from our M$^3$-20M dataset with their SMILES strings, 3D coordinates, and textual descriptions to the LLMs. The single-modal examples use only SMILES strings. 100 molecules per configuration are generated for performance assessment. We train MoFlow and D2L-OMP using QM9-MM on RTX4090 with 24GB memory with the batch size being 32. MoFlow is trained with 200 epochs, while D2L-OMP is trained with 300 epochs. The other parameters of these two models are the same as those in the original paper. We perform each experiment five times to ensure statistical reliability and report the averaged results and variance. 

\subsubsection{Performance Metrics}
We analyze the performance of models trained on single-modal and multi-modal datasets in molecule generation using three metrics: \textbf{validity}, \textbf{uniqueness}, and \textbf{novelty}. 

\textbf{Validity} measures the proportion of chemically correct molecular structures generated.
\textbf{Uniqueness} reflects the diversity of generated molecules.
\textbf{Novelty} is defined as the fraction of valid molecules that are not present in the training dataset, revealing the model’s ability to propose new molecular structures.  

Furthermore, we validate our results using the comprehensive MOSES benchmark with additional structural metrics. \textbf{Similarity to the nearest neighbor (SNN/Test)} measures reference manifold matching precision. \textbf{Fragment similarity (Frag/Test)} assesses BRICS fragment~\cite{degen2008art} distribution alignment in generated and reference sets. \textbf{Scaffold similarity (Scaff/Test)} evaluates Bemis-Murcko scaffold~\cite{bemis1996properties} distribution correspondence in generated and reference sets. \textbf{Filters} is the fraction of generated molecules that pass custom medicinal chemistry filters (MCFs) and PAINS filters~\cite{baell2010new}.

The MOSES benchmark includes key chemical property metrics (\textbf{logP}, \textbf{SA}, \textbf{QED}, \textbf{Molecular weight (MW)}). We calculate Wasserstein-1 distances between generated and test set distributions for these properties. Lower distances indicate better alignment with realistic molecular characteristics, further confirming generation quality.

\subsubsection{Experimental Results}
The results in Table \ref{tab:generation} demonstrate that models leveraging multi-modal datasets consistently outperform their single-modal counterparts using single-modal datasets across all metrics and model types.

For validity, both specialized molecule generation models and general LLMs show remarkable performance improvement with multi-modal data. 

For uniqueness, the consistently high uniqueness scores across all models demonstrate multi-modal datasets boost their robust capability to generate diverse molecular structures.

For novelty, our multi-modal dataset significantly enhances the performance of both LLMs and specialized molecule generation models. 

\begin{table*}[h!]
    \centering
        \caption{Molecule generation performance comparison: multi-modal data vs. single-modal data.}
    \label{tab:generation}
    \begin{tabular}{lcccccc}
        \toprule
        & \multicolumn{2}{c}{Validity(\%)↑} & \multicolumn{2}{c}{Uniqueness(\%)↑} & \multicolumn{2}{c}{Novelty(\%)↑} \\
        \cmidrule(r){2-3} \cmidrule(r){4-5} \cmidrule(r){6-7}
        & Single-modal & Multi-modal & Single-modal & Multi-modal & Single-modal & Multi-modal \\
        \midrule
         \multicolumn{7}{l}{\textbf{Specialized Molecule Generation Models}} \\
         MoFlow & 95.36($\pm$1.2) &  \textbf{96.21}($\pm$1.24) & 98.65($\pm$0.6) &  \textbf{98.71}($\pm$0.19) & 94.72($\pm$0.8) &  \textbf{99.12}($\pm$0.13) \\
        D2L-OMP & 98.60 ($\pm$0.83) &  \textbf{98.88}($\pm$0.92) & 99.80($\pm$0.17) &  \textbf{99.80}($\pm$0.14) & 83.39($\pm$0.52) &  \textbf{89.10}($\pm$0.51) \\
        \midrule
        \multicolumn{7}{l}{\textbf{General LLMs}} \\
        GLM4 & 72.73($\pm$1.73) & \textbf{85.37}($\pm$1.58) & 68.17($\pm$2.94) & \textbf{82.25}($\pm$2.95) & 97.90($\pm$0.98) & \textbf{98.10}($\pm$1.23) \\
        GPT-3.5 & 76.01($\pm$5.63) &  \textbf{84.8}($\pm$0.6) & 46.46($\pm$3.23) &  \textbf{93.86}($\pm$1.49) & 85.72($\pm$0.67) &  \textbf{96.51}($\pm$1.57) \\
        GPT-4 & 92.3($\pm$3.85) &  \textbf{97.99}($\pm$0.41) & 58.64($\pm$0.36) &  \textbf{70.17}($\pm$4.92) & 90.66($\pm$1.14) &  \textbf{98.95}($\pm$0.34) \\
        \bottomrule
    \end{tabular}%
\end{table*}

Furthermore, Table~\ref{tab:moses} summarizes the results of additional molecule generation performance metrics. Multi-modal training consistently enhances performance across nearly all indicators, enabling the model to exceed single-modal baselines in generating chemically meaningful structures. Improvements are evident in SNN, Frag, Scaff, and Filter scores. Multi-modal training also reduces Wasserstein-1 distance for logP, SA, and QED properties. These results demonstrate that the chemical properties of generated molecules align with those of the reference molecules, highlighting the effectiveness of multi-modal data in producing compounds with pharmacologically relevant characteristics while maintaining structural diversity and synthetic accessibility. 




\begin{table*}[h]
\centering
\caption{Molecule generation comparison using MOSES benchmark metrics. The table presents results of structural validity (SNN/Test, Frag/Test, Scaff/Test, Filters) and chemical property alignment (Wasserstein-1 distances for logP, SA, QED, MW) between using single-modal data and using multi-modal data. Lower Wasserstein-1 distance indicates better distribution matching with the reference compounds.}
\label{tab:moses}
\begin{adjustbox}{max width=1\textwidth}
\begin{tabular}{l l c c c}
\toprule
 & & \textbf{GLM4} & \textbf{GPT-3.5} & \textbf{GPT-4} \\
\midrule
\multirow{2}{*}{\textbf{SNN/Test(↑)}} & Single-modal & 0.49(±0.03) & 0.53(±0.09) & 0.49(±0.07) \\
 & Multi-modal & \textbf{0.49}(±0.09) & \textbf{0.56}(±0.04) & \textbf{0.52}(±0.01) \\
\midrule
\multirow{2}{*}{\textbf{Frag/Test(↑)}} & Single-modal & 0.24(±0.02) & 0.32(±0.12) & 0.19(±0.08) \\
 & Multi-modal & \textbf{0.27}(±0.14) & \textbf{0.44}(±0.01) & \textbf{0.20}(±0.01) \\
\midrule
\multirow{2}{*}{\textbf{Scaf/Test(↑)}} & Single-modal & 0.04(±0.05) & 6.90e-4(±6.90e-4) & 2.01e-3(±4.13e-4) \\
 & Multi-modal & \textbf{0.07}(±0.09) & \textbf{0.01}(±0.01) & \textbf{0.01}(±4.99e-3) \\
\midrule
\multirow{2}{*}{\textbf{Filters(↑)}} & Single-modal & 0.73(±0.09) & 0.8(±0.0) & 1.0(±0.0) \\
 & Multi-modal & \textbf{0.80}(±1.11e-16) & \textbf{1.0(±0.0)} & \textbf{1.0}(±0.10) \\
\midrule
\multirow{2}{*}{\textbf{logP(↓)}} & Single-modal & 1.35(±0.17) & 1.63(±0.32) & 1.27(±0.31) \\
 & Multi-modal & 1.44(±0.28) & \textbf{0.99(±0.05)} & \textbf{0.92(±0.11)} \\
\midrule
\multirow{2}{*}{\textbf{SA(↓)}} & Single-modal & 0.83(±0.22) & 0.63(±0.11) & 0.60(±0.14) \\
 & Multi-modal & 1.00(±0.10) & \textbf{0.57(±0.02)} & \textbf{0.54}(±0.06) \\
\midrule
\multirow{2}{*}{\textbf{QED(↓)}} & Single-modal & 0.20(±0.04) & 0.15(±0.05) & 0.15(±0.06) \\
 & Multi-modal & \textbf{0.20(±0.08)} & \textbf{0.12}(±0.03) & \textbf{0.12}(±0.02) \\
\midrule
\multirow{2}{*}{\textbf{Molecular weight(↓)}} & Single-modal & 112.10(±20.66) & 67.83(±15.92) & 89.75(±37.98) \\
 & Multi-modal & \textbf{105.72}(±38.35) & 68.01(±4.04) & 95.62(±9.33) \\
\bottomrule
\end{tabular}
\end{adjustbox}
\end{table*}

\begin{table*}[h]
\centering
\caption{Model performance comparison among different datasets.}
\label{tab:datasets}
\begin{tabular}{>{\centering\arraybackslash}m{3cm} >{\centering\arraybackslash}m{3cm} >{\centering\arraybackslash}m{3cm} >{\centering\arraybackslash}m{3cm}}
\toprule
 Datasets& Validity(\%)↑ & Uniqueness(\%)↑ & Novelty(\%)↑ \\
\midrule
QM9~\cite{ramakrishnan2014quantum} & 75.17 (±1.43) & 64.13 (±1.58) & 89.43 (±0.01) \\
Pcdes~\cite{zeng2022deep} & 84.31 (±0.24) & 82.64 (±1.60) & 93.253 (±0.59) \\
PubChemSTM~\cite{liu2023multi} & 82.24 (±0.70) & 80.89 (±1.12) & 95.91 (±1.93) \\
igcdata~\cite{liu2024git} & 78.34 (±2.22) & 61.21 (±1.50) & 95.07 (±0.07) \\
Ours & \textbf{84.8} (±0.6) & \textbf{93.86} (±1.49) & \textbf{96.51} (±1.57) \\
\bottomrule
\end{tabular}%
\end{table*}

Finally, we compare the molecule generation performance of GPT-3.5 using various datasets—QM9, Pcdes, PubChemSTM, igcdata, and our M$^3$-20M—focusing on three key metrics: Validity, Uniqueness, and Novelty (see Table~\ref{tab:datasets}). The results demonstrate the advantages of our M$^3$-20M dataset.

For validity, GPT-3.5 consistently generates the most chemically accurate molecules using our dataset, underscoring M$^3$-20M's effectiveness in capturing essential chemical properties. Regarding uniqueness, the molecules produced with our data exhibit a greater diversity of structures, indicating the potential for discovering novel compounds. This trend continues with the novelty metric, that is, using our dataset can generate molecular structures distinct from those seen during training, further validating its ability to foster innovation in molecular design. 

Overall, the results above indicate that our dataset can significantly benefit molecule generation,  offering a valuable resource for AI-driven molecule design and discovery. 

\subsection{Molecular Property Prediction Experiments}
Accurate prediction of molecular properties is fundamental to drug discovery. To evaluate M$^3$-20M's effectiveness in property prediction tasks, we conduct extensive experiments following the settings in MoleculeNet~\cite{wu2018moleculenet}. Our evaluation encompasses regression and classification tasks, addressing diverse molecular properties crucial for drug development. We employ FP-ICL for regression tasks, both FP-ICL and parameter-efficient fine-tuning with LoRA for classification tasks. To prevent data leakage, we firstly partition datasets into non-overlapping training and test sets. Secondly, we mask all the target property-related information in textual descriptions during training and evaluation. We perform each experiment five times and report the averaged result and variance. 

\subsubsection{Regression Tasks}

\textbf{Experimental details}.  
We apply FP-ICL for regression tasks on the QM9-MM dataset under three settings: using SMILES strings only, using both SMILES strings and 3D coordinates, and using SMILES strings with 3D coordinates plus textual descriptions. We focus on predicting dipole moment ($\mu$) and isotropic polarizability ($\alpha$). We use GPT-3.5, GPT-4, and GLM-4 APIs as LLMs, and randomly select four molecules from QM9-MM for the LLMs in the prompt according to the three specific settings.

\textbf{Performance metrics}. 
The metric we use for the regression tasks is Mean Absolute Error (MAE), measuring the average error between predicted values and actual values. The smaller the MAE value, the more accurate the prediction. Formally, MAE is evaluated as follows:
\begin{equation}
\text{MAE} = \frac{1}{n} \sum_{i=1}^{n} | y_i - \hat{y}_i |
\end{equation}
where $n$ is the number of samples, $y_i$ is predicted value, and $\hat{y}_i$ is the ground truth value.

\textbf{Experimental results}. 
The results in Table~\ref{tab:comparison} show that providing more information with multi-modal data in the prompt can help LLMs better understand molecules and their chemical properties. Except for the task of predicting the dipole moment using GPT-4, all other models and tasks achieve the best results on the multi-modal dataset, which proves the effectiveness and high quality of our dataset.

\begin{table*}[ht]
\centering
\caption{Comparison of molecular property prediction performance using multi-modal and single-modal data (regression task).}
\label{tab:comparison}
\begin{tabular}{>{\centering\arraybackslash}m{1.8cm} >{\centering\arraybackslash}m{1.8cm} >{\centering\arraybackslash}m{2cm} >{\centering\arraybackslash}m{1.8cm} >{\centering\arraybackslash}m{2cm} >{\centering\arraybackslash}m{1.8cm} >{\centering\arraybackslash}m{2cm}}
\toprule
 & \multicolumn{2}{c}{SMILES} & \multicolumn{2}{c}{SMILES+3D} & \multicolumn{2}{c}{SMILES+3D+texts} \\
\cmidrule(lr){2-3} \cmidrule(lr){4-5} \cmidrule(lr){6-7}
 & $\mu$ & $\alpha$ & $\mu$ & $\alpha$ & $\mu$ & $\alpha$ \\
\midrule
GPT-3.5 & 1.78 (±0.16) & 39.50 (±2.16) & 1.82 (±0.36) & 35.37 (±1.53) & \textbf{1.56} (±0.28) & \textbf{34.74} (±1.60) \\
GLM-4 & 1.27 (±0.13) & 12.82 (±0.77) & 1.13 (±0.07) & 14.53 (±0.67) & \textbf{1.01} (±0.13) & \textbf{12.05} (±0.48) \\
GPT-4 & \textbf{1.17} (±0.12) & 13.71 (±0.08) & 1.28 (±0.20) & 28.9 (±0.84) & 1.27 (±0.30) & \textbf{9.50} (±1.65) \\
\bottomrule
\end{tabular}%
\end{table*}

\subsubsection{Classification Tasks}
\textbf{Experimental details}. 
First, we apply FP-ICL to regression tasks on BACE-MM, BBBP-MM, ClinTox-MM, HIV-MM, and Tox21-MM datasets with single-modal and multi-modal data respectively. Four randomly-selected molecules from each of the above datasets are used, with SMILES strings only for single-modal molecules, and SMILES strings, 3D coordinates and textual descriptions for multi-modal molecules.

Then, we employ parameter-efficient fine-tuning with LoRA on Llama3-8b using our dataset M$^3$-20M. Experiments are run on an RTX4090 with 24GB memory. LoRA hyperparameters include a rank of 4 and alpha of 16. Training lasts a maximum of 200 epochs with early stopping after 15-30 epochs without improvement.
 
\textbf{Performance metrics}. 
We perform each experiment five times and report the averaged result and variance,
using three metrics including accuracy mean (ACC Mean), accuracy variance (ACC Variance), and accuracy standard deviation (ACC Standard Deviation). These metrics provide comprehensive insight into prediction reliability and consistency across the different molecule datasets.



\textbf{Experimental results}.  
Results from FP-ICL on classification tasks (Table~\ref{tab:MPP}) show using multi-modal data generally outperforms using single-modal data in molecular property prediction. On the majority of datasets including BACE-MM, BBBP-MM and ClinTox-MM, mean accuracy is improved,  only on HIV-MM there is a slight performance decrease. Using multi-modal data usually yields reduced variance and standard deviation, particularly on the ClinTox-MM dataset, indicating enhanced prediction stability and consistency. These findings suggest that integrating multiple modalities provides complementary information and leads to more robust and accurate molecular property predictions.

\begin{table*}[t]
\centering
\caption{Comparison of molecular property prediction performance using multi-modal and single-modal data (classification task).}
\label{tab:MPP}
\small 
\begin{tabular}{>{\centering\arraybackslash}m{2.2cm} >{\centering\arraybackslash}m{1cm} >{\centering\arraybackslash}m{1cm} >{\centering\arraybackslash}m{1cm} >{\centering\arraybackslash}m{1cm} >{\centering\arraybackslash}m{1cm} >{\centering\arraybackslash}m{1cm}}
\toprule
 & \multicolumn{2}{c}{ACC Mean} & \multicolumn{2}{c}{ACC Variance} & \multicolumn{2}{c}{ACC Standard Deviation} \\
\cmidrule(lr){2-3} \cmidrule(lr){4-5} \cmidrule(lr){6-7}
 & Single-modal & Multi-modal & Single-modal & Multi-modal & Single-modal & Multi-modal \\
\midrule
BACE-MM & 0.5680 & \textbf{0.5780} & 0.0014 & 0.0018 & 0.0377 & 0.0427 \\
BBBP-MM & 0.2280 & \textbf{0.2720} & 0.0005 & 0.0008 & 0.0217 & 0.0277 \\
ClinTox-MM & 0.9260 & \textbf{0.9280} & 0.0005 & 0.0002 & 0.0230 & 0.0130 \\
HIV-MM & \textbf{0.9740} & 0.9680 & 0.0002 & 0.0002 & 0.0134 & 0.0148 \\
\bottomrule
\end{tabular}%
\end{table*}

Results on the Tox21-MM dataset are summarized in Table~\ref{tab:Tox21}, from which we can still see that  using multi-modal data can generally enhance molecular property prediction performance across multiple toxicity subtasks. On 8 of 12 subtasks we witness improved or equivalent accuracy, with notable improvements in NR-AhR and NR-Aromatase tasks. Using multi-modal data also frequently produces more consistent predictions, evidenced by the reduced variance and standard deviation in several subtasks, particularly NR-ER and SR-HSE. These results confirm that complementary information from multiple modalities enhances accuracy and reliability of toxicity prediction. 

\begin{table*}[t]
\centering
\caption{Comparison of molecular property prediction performance using multi-modal and single-modal data on Tox21-MM (classification task).}
\label{tab:Tox21}
\small 
\begin{tabular}{>{\centering\arraybackslash}m{2cm} >{\centering\arraybackslash}m{1.5cm} >{\centering\arraybackslash}m{1.5cm} >{\centering\arraybackslash}m{1.5cm} >{\centering\arraybackslash}m{1.5cm} >{\centering\arraybackslash}m{1.5cm} >{\centering\arraybackslash}m{1.5cm}}
\toprule
 & \multicolumn{2}{c}{ACC Mean} & \multicolumn{2}{c}{ACC Variance} & \multicolumn{2}{c}{ACC Standard Deviation} \\
\cmidrule(lr){2-3} \cmidrule(lr){4-5} \cmidrule(lr){6-7}
 & Single-modal & Multi-modal & Single-modal & Multi-modal & Single-modal & Multi-modal \\
\midrule
NR-AR & 0.9620 & \textbf{0.9680} & 0.0003 & 0.0003 & 0.0164 & 0.0164 \\
NR-AR-LBD & 0.9800 & \textbf{0.9800} & 0.0002 & 0.0003 & 0.0141 & 0.0173 \\
NR-AhR & 0.8740 & \textbf{0.9020} & 0.0006 & 0.0009 & 0.0251 & 0.0303 \\
NR-Aromatase & 0.9620 & \textbf{0.9720} & 0.0001 & 0.0005 & 0.0110 & 0.0217 \\
NR-ER & 0.9140 & \textbf{0.9240} & 0.0014 & 0.0003 & 0.0378 & 0.0167 \\
NR-ER-LBD & \textbf{0.9580} & 0.9560 & 0.0003 & 0.0009 & 0.0179 & 0.0297 \\
NR-PPAR-gamma & 0.9780 & \textbf{0.9780} & 0.0002 & 0.0002 & 0.0148 & 0.0148 \\
SR-ARE & \textbf{0.9060} & 0.8680 & 0.0002 & 0.0005 & 0.0152 & 0.0217 \\
SR-ATAD5 & \textbf{0.9680} & 0.9600 & 0.0005 & 0.0003 & 0.0228 & 0.0158 \\
SR-HSE & 0.9520 & \textbf{0.9560} & 0.0009 & 0.0001 & 0.0295 & 0.0114 \\
SR-MMP & 0.8820 & \textbf{0.8960} & 0.0008 & 0.0016 & 0.0277 & 0.0404 \\
SR-p53 & \textbf{0.9500} & 0.9340 & 0.0005 & 0.0014 & 0.0235 & 0.0378 \\
\bottomrule
\end{tabular}%
\end{table*}

Table~\ref{tab:llama3} presents the results of fine-tuning Llama3-8b with our M$^3$-20M dataset. We also see notable improvements on model performance across various tasks. Concretely, using multi-modal data achieves significant gain over using single-modal data on BACE-MM, Tox21-MM, HIV-MM, and ClinTox-MM datasets,  with the most pronounced improvement on the BACE dataset, where performance increases from 0.844 to 0.887. This shows that our dataset's value and potential. Our dataset can also be used for other downstream tasks such as name prediction, reaction prediction, retrosynthesis, text-based molecule design, and molecule captioning, etc.

\begin{table}[ht]
    \centering
    \caption{Fine-tuning Llama3-8b with our M$^3$-20M dataset demonstrates significant performance improvement in terns of ACC mean.}
    \label{tab:llama3}
    \begin{tabular}{@{}lcc@{}}
        \toprule
        \textbf{Dataset} & \textbf{Single-modal} & \textbf{Multi-modal (ours)} \\ 
        \midrule
        BBBP-MM      & 0.959 & \textbf{0.959} \\ 
        BACE-MM      & 0.844 & \textbf{0.887} \\ 
        Tox21-MM     & 0.817 & \textbf{0.849} \\ 
        HIV-MM       & 0.780 & \textbf{0.803} \\ 
        ClinTox-MM   & 0.985 & \textbf{0.998} \\ 
        \bottomrule
    \end{tabular}
\end{table}

\subsection{Statistic Analyses and Results}
This section presents statistical information on the M$^3$-20M dataset to highlight its comprehensive annotations, diverse textual descriptions, and physicochemical properties, which can significantly enhance its utility for molecular research and AI-driven drug discovery. 

\subsubsection{Textual Descriptions} 
The PubChem description bar is relatively small, reaching only 360,133, while the M$^3$-20M annotation description bar towers over it, marking a count of 20,249,090. This enhancement is critical for AI-driven drug design and discovery, as the increased volume and details of descriptions facilitate more comprehensive analysis and model training.


\subsubsection{Word Distribution in Description Texts} 
\begin{figure}[h]
\centering
{\includegraphics[width=0.45\textwidth]{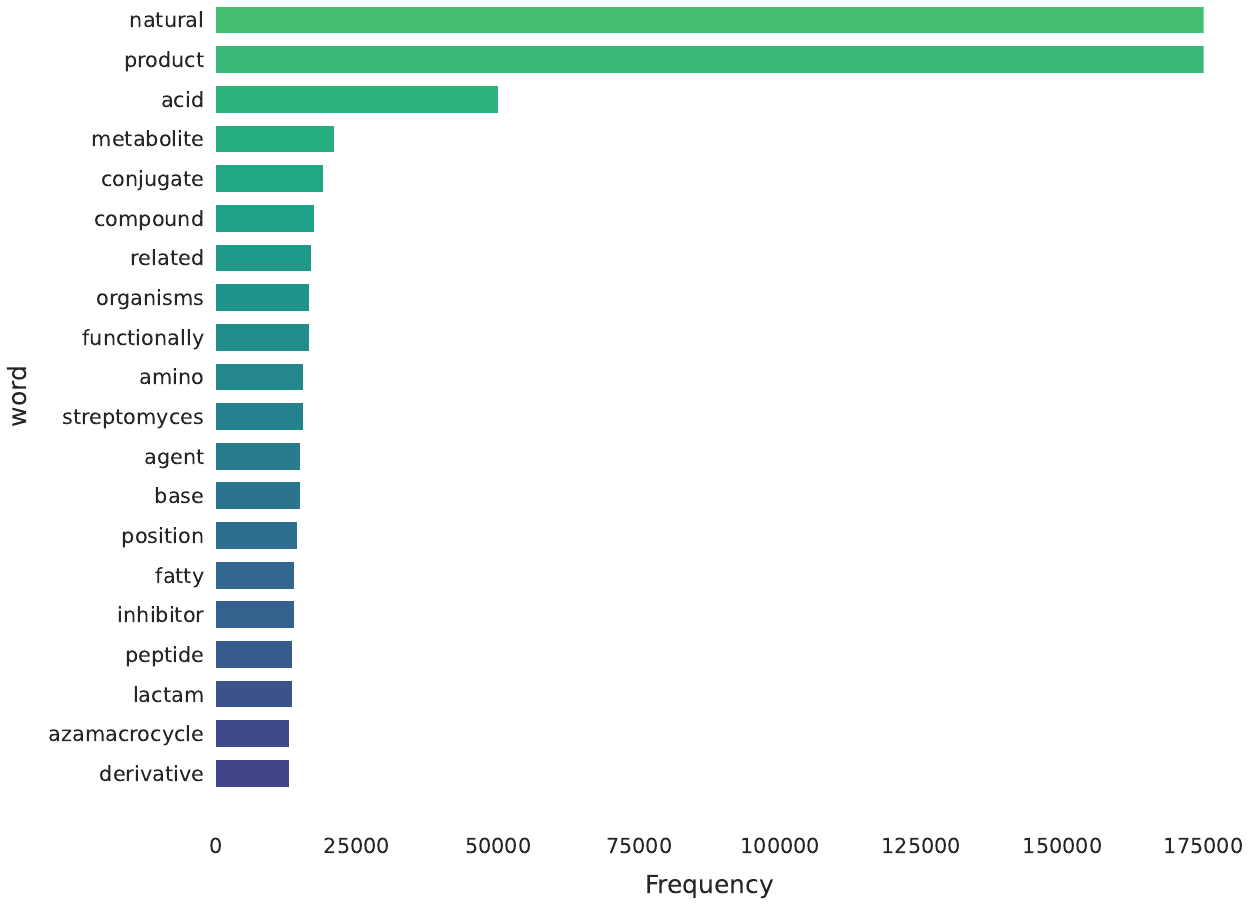}}
\caption{20 most-frequent words in textual descriptions.}
\label{fig:words}
\end{figure}

Figure~\ref{fig:words} shows the 20 most frequently occurring words in textual descriptions of our dataset, including ``natural'',  ``product'', ``acid'', and ``metabolite'' etc. These terms appear tens of thousands of times, reflecting the dataset's emphasis on biologically derived compounds. The frequency distribution of descriptors like ``conjugate'',  ``streptomyces'' and ``aza macrocycle'' reveals the dataset's focus on microbial secondary metabolites and their derivatives. This lexical analysis provides insight into the composition of the textual descriptions and highlights the predominance of natural product chemistry within the represented molecular space. 

\begin{table*}[t]
\centering
\caption{The 26 key physicochemical properties and their corresponding numbers and references.}
\label{tab:chemical_physical_properties}
\begin{tabular}{>{\raggedright\arraybackslash}m{6cm} >{\centering\arraybackslash}m{2cm} >{\centering\arraybackslash}m{6cm}}
\toprule
\textbf{Physicochemical Properties} & \textbf{Number} & \textbf{Reference} \\
\midrule
Molecular Weight & 19174000 & Computed by PubChem 2.2 \\
XLogP3 & 16340203 & Computed by XLogP3 3.0 \\
Hydrogen Bond Donor Count & 19173687 & Computed by Cactvs 3.4.8.18 \\
Hydrogen Bond Acceptor Count & 19173559 & Computed by Cactvs 3.4.8.18 \\
Rotatable Bond Count & 19173464 & Computed by Cactvs 3.4.8.18 \\
Exact Mass & 19173398 & Computed by PubChem 2.2 \\
Monoisotopic Mass & 19173327 & Computed by PubChem 2.2 \\
Topological Polar Surface Area & 19173287 & Computed by Cactvs 3.4.8.18 \\
Heavy Atom Count & 19173259 & Computed by PubChem \\
Formal Charge & 19173222 & Computed by PubChem \\
Complexity & 19173210 & Computed by Cactvs 3.4.8.18 \\
Isotope Atom Count & 19173193 & Computed by PubChem \\
Defined Atom Stereocenter Count & 19173170 & Computed by PubChem \\
Undefined Atom Stereocenter Count & 19173153 & Computed by PubChem \\
Defined Bond Stereocenter Count & 19173147 & Computed by PubChem \\
Undefined Bond Stereocenter Count & 19173137 & Computed by PubChem \\
Covalently-Bonded Unit Count & 19173132 & Computed by PubChem \\
Compound Is Canonicalized & 19173130 & Computed by PubChem\\
Physical Description & 2031 & CAMEO Chemicals database \\
Color/Form & 246 & Hazardous Substances Data Bank (HSDB) \\
Odor & 100 & Hazardous Substances Data Bank (HSDB) \\
Boiling Point & 130 & CAMEO Chemicals database\\
Melting Point & 633 & CAMEO Chemicals database \\
Flash Point & 35 & CAMEO Chemicals database\\
Solubility & 10813 & CAMEO Chemicals database\\
Density & 170 & CAMEO Chemicals database\\
\bottomrule
\end{tabular}
\end{table*}



\begin{figure}[h]
\centering
\includegraphics[width=0.4\textwidth]{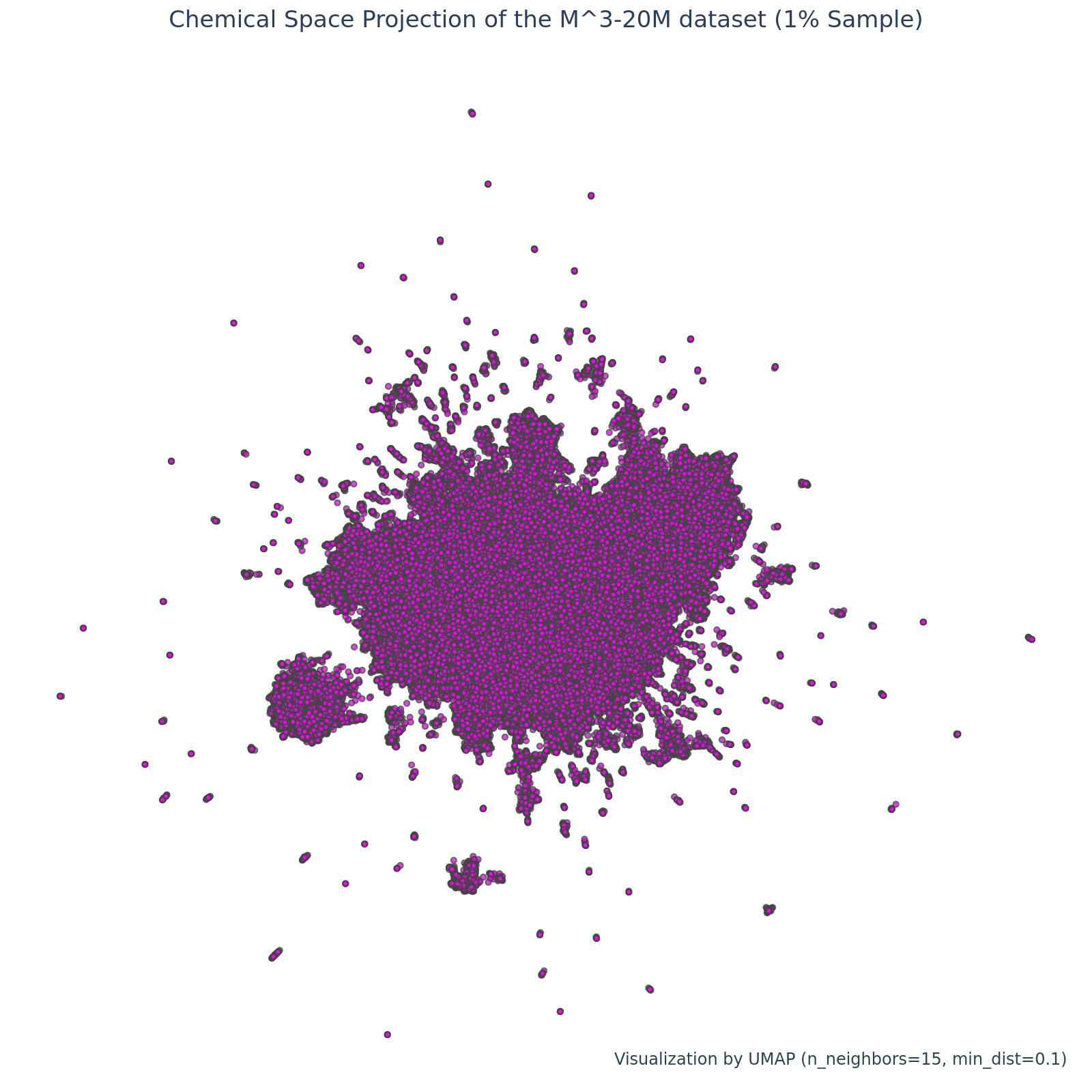}
\caption{Visualization of molecular space with uniform manifold approximation and projection (UMAP).}
\label{fig:umap}
\end{figure}


Figure~\ref{fig:umap} illustrates the diversity of our dataset in a two-dimensional space using UMAP~\cite{mcinnes2018umap}. UMAP (Uniform Manifold Approximation and Projection) excels as a dimensionality reduction technique by preserving both local and global structure while offering computational efficiency. It is particularly valuable for visualizing high-dimensional molecular fingerprints where topological relationships between compounds must be maintained. The UMAP projection of the M$^3$-20M dataset reveals a predominantly interconnected chemical space with a dense central region and several peripheral clusters, suggesting substantial molecular diversity within a coherent chemical landscape.  

\subsubsection{Dataset Maintenance}
A six-member maintenance team (three doctoral and three master's students) will perform bi-weekly updates to the M$^3$-20M dataset, correcting errors and implementing updates. This protocol enhances the dataset's reliability for molecular representation research and supports more robust model training.

\section{Conclusion}
In this paper, we introduce M$^3$-20M, a new and large-scale multi-modal molecule dataset containing over 20 million molecules. Extensive experiments demonstrate M$^3$-20M's effectiveness in boosting the performance of various models in molecule generation and property prediction. M$^3$-20M's integration of SMILES strings, 2D graphs, 3D structural data, physicochemical properties, and molecular textual descriptions provides a rich source for developing high-performance generative models for AI-driven drug design and discovery.


\section{Data and Software Availability Statement}
The dataset is openly accessible under the GNU General Public License v3.0, hosted on a publicly available platform \url{(https://github.com/bz99bz/M-3)}. 


\section{Author contributions statement}
Siyuan Guo and Shuigeng Zhou conceived the experiment(s), Siyuan Guo, Lexuan Wang, Chang Jin, Jinxian Wang, Han Peng, and Huayang Shi conducted the experiment(s), Siyuan Guo, Lexuan Wang, and Chang Jin analysed the results. Siyuan Guo, Wengen Li, Jihong Guan, and Shuigeng Zhou wrote and reviewed the manuscript.

\section{Acknowledgments}
The authors thank the anonymous reviewers for their valuable suggestions. This work was supported in part by the National Natural Science Foundation of China (NO. 62372326, No. 62172300).

\begin{tcolorbox}[colframe=black, colback=white, sharp corners]
\textbf{Key Points}
\begin{itemize}
\item M$^3$-20M is the largest open-access multi-modal molecule dataset for AI-driven drug design and discovery, with over 20 million molecules.
\item M$^3$-20M has comprehensive modalities, integrating SMILES, 2D/3D structures, physicochemical properties, and textual descriptions.
\item Experiments show that M$^3$-20M can significantly boost the model performance in molecular generation and property prediction tasks.
\item M$^3$-20M supports diverse downstream tasks, including molecule generation, molecular property prediction, lead optimization, and virtual screening, etc.
\end{itemize}
\end{tcolorbox}

\bibliographystyle{plain}
\bibliography{reference}

\end{document}